\documentclass[aps,groupedaddress,showpacs,letterpaper,twocolumn]{revtex4}
\usepackage{graphicx} 
\usepackage{amsmath}

\begin{document}

\title{Entangling single and $N$ atom qubits for fast quantum state detection and transmission }
\author{ M. Saffman and T. G. Walker}
\affiliation{
Department of Physics,
University of Wisconsin, 1150 University Avenue,  Madison, Wisconsin 53706
}
 \date{\today}

\begin{abstract}
We discuss the use of Rydberg blockade techniques for
entanglement of $1$ atom qubits with collective $N$ atom qubits. We show how the entanglement can be used to achieve fast readout and transmission of the state of single atom qubits without the use of optical cavities. 
\end{abstract}

\maketitle

\section{Introduction}

Isolated neutral atoms trapped by electromagnetic fields are being studied intensively as a way of implementing a scalable quantum processor. 
In this approach to quantum logic the qubit basis states are represented by different 
ground state hyperfine levels. The hyperfine states have decoherence times that are potentially 
many seconds long, limited mainly by weak interactions with the trapping fields and background gas collisions due to imperfect vacuum conditions. Despite their superb isolation from the environment the qubits can be rapidly initialized and  manipulated with near resonant optical fields through optical pumping and stimulated Raman processes. A number of protocols for two-qubit gates have been proposed
including ground state collisions\cite{ref.collision}, optically induced short range dipole-dipole interactions\cite{deutsch2}, and dipole-dipole interactions between highly excited Rydberg states\cite{ref.blockade1,ref.blockade2}.
Among the different approaches to implementing neutral atom logic gates the Rydberg gate is unique in that the gate operation is essentially independent of the motional state of the atoms. Because of this the Rydberg gate is not limited to speeds that are slow compared to trap vibrational frequencies and can potentially be run at MHz rates or faster\cite{ref.swqclong}.   

The combination of long decoherence times and fast gate operation is advantageous for building a scalable quantum processor. However, the requirements of fast state readout and state transmission in a quantum network environment are not easily satisfied. In this work we examine
coherent coupling of a single atom to a many atom mesoscopic qubit for enhanced measurement and transmission capabilities.  The approach described here may be generic
in the sense that introduction of a mesoscopic intermediary helps to ``impedance match" the flow of information from the 
microdomain of a single atom  to the macroscopic propagation distance available to a single photon. 
The approach we describe complements recent research that has resulted in the capability of storing and generating photonic states using mesoscopic atomic ensembles\cite{refs.ensembleexperiment,ref.kimblerepeater} and may lead to the creation of distant entanglement between ensembles\cite{polzik1}.

It may be preferable to store information in single atom qubits, as opposed to ensembles, since they have substantially longer coherence times due to the absence of collisions.  
State measurement on atomic qubits can be performed by measuring optically induced single atom fluorescence\cite{ref.winelandshelving}. Alternatively  amplitude\cite{ref.wineland}  and/or phase shifts imparted to a tightly focused probe beam can be used. In either case Poissonian photon counting statistics result in measurement times that are several orders of magnitude longer than Rydberg gate operation times. While the use of subPoissonian light could be advantageous in this context, it would add additional complexity.    Quantum state transmission between single atoms over a photonic channel requires strong atom to photon coupling and high fidelity protocols for state transmission and distant entanglement\cite{cirac,lloyd}  are based on interfacing atoms with  ultra high finesse  optical cavities. While impressive experimental advances have been made in achieving strong coupling of a single atom to a cavity mode\cite{cqed} the problem of coupling a large number of atomic qubits in a quantum processor to high finesse cavities remains an extremely demanding challenge.  

\begin{figure}[!t]
\begin{minipage}[c]{8.5cm}
\includegraphics[width=8.5cm]{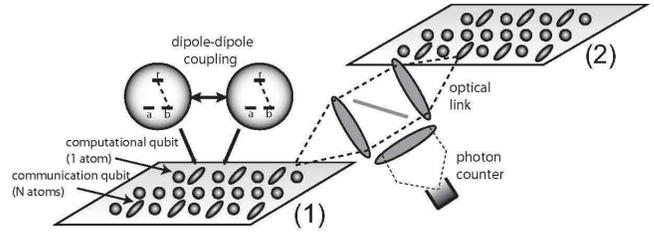}
\caption{Heterogeneous arrays of single atom and collective qubits interconnected via a photonic channel.   }
\label{fig.qubits}
\end{minipage}
\end{figure}

In this work we explore a new paradigm for fast single atom state detection, and transmission between distant qubits 
without the use of optical cavities, as shown in Fig. \ref{fig.qubits}. In Sec. \ref{sec.readout} we describe a method for fast state readout based on coupling single atom qubits to $N$ atom collective qubits\cite{ref.blockade2}.  
A protocol for state  transmission is given in Sec. \ref{sec.transmission} and its expected fidelity is calculated.

\section{Fast state measurement}
\label{sec.readout}

\begin{figure}[!t]
\begin{minipage}[c]{7.5cm}
\includegraphics[width=7.5cm]{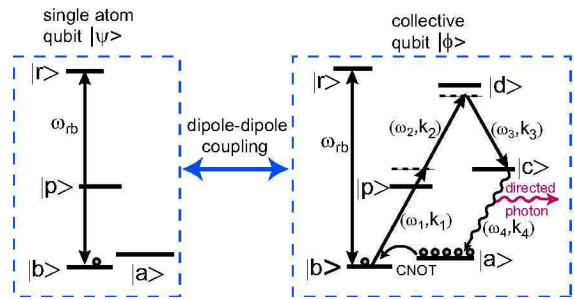}
\caption{(color online) Atomic levels for interfacing a single atom qubit to a collective qubit that drives a directed photon emission. A time reversed version maps an incoming photon onto a qubit state. See the text for details. }
\label{fig.levels}
\end{minipage}
\end{figure}

Reliable measurement of the quantum state of a single atom using light scattering requires a finite integration time due to Poissonian photon statistics. Numerical estimates for free space\cite{ref.swqclong} as well as 
microcavity geometries\cite{ref.schmiedmayeratomchip} suggest that integration times of a few tens of $\mu\rm s$ are necessary for the measurement error to be small while recent ion trap experiments have demonstrated low measurement errors with a few hundred $\mu\rm s$ of integration time\cite{ref.winelandenhancedmeasurement}. 
These times are one to two orders of magnitude longer than projected gate speeds using Rydberg atoms\cite{ref.swqclong}. 
In order to  achieve readout and transmission rates of much higher speed and fidelity than are possible with individual atoms we propose to use cross entanglement between  a single atom with quantum state $|\psi\rangle$ and  a spatially separated $N$ atom collective qubit with quantum state $|\phi\rangle$ as pictured in Fig. \ref{fig.levels}. 

 The single atom qubit state is $|\psi\rangle=c_a|a\rangle + c_b|b\rangle$ where $|a\rangle,|b\rangle$ are the hyperfine ground states shown in Fig. \ref{fig.levels} and $c_a,c_b$ are c-numbers. A collective $N$ atom qubit state can be written as 
$|\phi\rangle=c_{\bar a}|\bar a\rangle + c_{\bar b}|\bar b\rangle$ where
$|\bar a\rangle=|a_1a_2...a_N\rangle$ and $|\bar b\rangle=\frac{1}{\sqrt N}
\sum_j|a_1a_2...b_j..a_N\rangle \equiv\frac{1}{\sqrt N} \sum_j |a_{\tilde \jmath};b_j\rangle$ with $c_{\bar a}, c_{\bar b}$ c-numbers. 
  Referring to Fig. \ref{fig.levels} we recall that the protocol for a two-qubit Rydberg phase gate between qubits $|\psi\rangle_1,|\psi\rangle_2$ 
is\cite{ref.blockade1}:  $\pi$ pulse on $|\psi\rangle_1$, 
$2\pi$ pulse on $|\psi\rangle_2$, 
$\pi$ pulse on $|\psi\rangle_1$, where the Rabi pulses connect 
levels $|b\rangle$ and $|r\rangle.$ In the limit of strong dipole-dipole interaction between Rydberg atoms the gate operates in a dipole-blockade mode, giving a high fidelity phase gate that is insensitive to the exact value of the dipole-dipole coupling. 
 The protocol for a collective two-qubit phase gate between  $|\phi\rangle_1,|\phi\rangle_2$ 
is identical. In terms of the actual implementation the main differences are that the pulse areas are $\sqrt N$ smaller for a  collective qubit and the Rydberg level must be 
 chosen with care to avoid  zeroes in the dipole-dipole 
interaction\cite{foster1,foster2}. 
Taking these differences into account it is apparent that we can also 
achieve a mixed gate between single atom and collective qubits from the sequence:  
$\pi$ pulse on $|\psi\rangle$, 
$2\pi$ pulse on $|\phi\rangle$, 
$\pi$ pulse on $|\psi\rangle$. Sandwiching this operation with single qubit rotations\cite{chuangbook} we obtain a controlled not (CNOT) gate 
$|\psi\rangle_1|\phi\rangle_2\rightarrow|\psi\rangle_1|\phi\oplus\psi\rangle_2$
between  the single atom qubit $|\psi\rangle_1$ and the collective qubit $|\phi\rangle_2.$

Having entangled $|\psi\rangle$  and $|\phi\rangle$ we can use the mesoscopic qubit to enhance the rate of measurement of the state of $|\psi\rangle.$ In order to achieve  rapid state readout by single atom fluorescence a large detection solid angle is required to collect the spontaneous photons emitted in random directions. Simply using a very fast lens is not a viable option in the presence of  conflicting experimental requirements. As discussed in\cite{ref.sw1} a mesoscopic 
qubit can be prepared in a state that leads to a spatially directed 
single photon emission pattern. We therefore propose to prepare
$|\phi\rangle$ to give a spatially directed emission that can be detected without requiring a large solid angle for the detection optics. 
This approach provides a speedup as long as the slowdown due to the  additional overhead of local operations is smaller than the speedup of the  photon detection rate due to the directional emission. As we discuss below this appears possible  for experimentally accessible parameters.

To create the spatially directed emission a three photon $\pi$ pulse connecting $|b\rangle$ with $|c\rangle$ using the fields $(\omega_1,{\bf k}_1),(\omega_2,{\bf k}_2),(\omega_3,{\bf k}_3)$  shown in  Fig. \ref{fig.levels} gives the mapping\cite{ref.sw1}
\begin{equation}
|\phi\rangle=\frac{1}{\sqrt N}\sum_j |a_{\tilde \jmath};b_j\rangle \rightarrow |\phi_e\rangle=\frac{1}{\sqrt N}\sum_j e^{\imath\varphi_j} |a_{\tilde \jmath};c_j\rangle
\label{eq.mapping}
\end{equation}
with $\varphi_j=(-{\bf k}_1-{\bf k}_2+{\bf k}_3)\cdot{\bf r}_j,$ 
 ${\bf r}_j$ is the position of atom $j$ and $|c\rangle$ is an excited  $P$ state connected to $|a\rangle$ by a closed transition.  Decay of the excited state $|\phi_e\rangle$ results in preferential emission of a single photon into the mode directed along the phase matched direction ${\bf k}_4={\bf k}_1+{\bf k}_2-{\bf k}_3.$

We note that since the entangled collective state is prepared by the CNOT operation the fields used to perform the mapping of Eq. (\ref{eq.mapping})
need not involve Rydberg levels. In particular, as shown in Fig. \ref{fig.levels}, we can choose to employ near resonance with a low lying auxiliary level $|d\rangle$  that has  an oscillator strength for a transition from the ground state that is large compared to that for transitions to  highly excited Rydberg levels.  The additional  level $|p\rangle$ is used for two-photon single qubit operations needed to complete the CNOT operation, and for enhancement of the three-photon $|b\rangle\rightarrow|c\rangle$ mapping. Provided there is a non-zero Clebsch-Gordan coefficient the  mapping $|b\rangle \rightarrow |c\rangle$ could also be achieved with a single photon transition, as in recent experiments on photon storage and generation with atomic ensembles\cite{refs.ensembleexperiment,ref.kimblerepeater}. In such a case the phase matching condition is simply ${\bf k}_4={\bf k}_1$ so the photon emission is collinear with the preparation pulse. With judicious choice of experimental parameters one polarization component of the emitted photon can still be separated from an orthogonally polarized excitation pulse. While this implies a  50 \% loss in efficiency it would simplify the experimental implementation.  
We concentrate here on a four photon scheme which has the advantage of allowing  the generated photon to be very efficiently separated from the excitation pulse due to its different direction of propagation.

The probability of emission into the mode along ${\bf k}_4$ can be calculated as follows. 
The $N$ atoms are assumed confined in an ellipsoidally shaped  harmonic trap that is created by  focusing  a single far off resonance laser beam with wavelength $\lambda_f$ to a Gaussian waist of radius $w_f.$ In a parabolic approximation, valid when the atomic temperature $T_a$ is much less than $T_f$ (the trap depth expressed in temperature units) the density distribution 
can be written as $n({\bf r})=n_0e^{-(\rho^2+ z^2/\xi^2)/w_a^2}.$
Here $\rho^2=x^2+y^2$ is the transverse coordinate, $z$ is along the symmetry axis of the trap,  $w_a=\sqrt{(3/2)T_{rel}}w_f$ is the width of the atomic cloud, $T_{rel}=T_{a}/T_f$, and 
the aspect ratio is $\xi=\pi w_f/\lambda_f.$ The
total number of atoms in the ellipsoidal volume is then 
$N= n_0 (3\pi T_{rel}/2)^{3/2}\xi w_f^3$ and the  
 radiated mode will  be of Gaussian form with a waist radius $w_0= \sqrt 2 w_{a}.$ 

This Gaussian mode radiates coherently into a solid angle $\Omega_c=2\pi/(k w_0)^2.$  The probability of phase-matched photon emission into this mode is 
$P_c\simeq \epsilon N \Omega_c $ and the probability of emission into a random direction outside of this mode is $P_i\simeq \epsilon (4\pi-\Omega_c)$ where $\epsilon$ is a normalization constant proportional to 
the emission probability per unit solid angle.  Requiring $P_c+P_i=1$ gives
\begin{equation}
P_c\simeq \frac{1}{1+\frac{2 k^2 w_0^2}{N}}=\frac{{\mathcal C}}{1+{\mathcal C}}.
\label{eq.pc}
\end{equation}
We see that when $N\gg k^2 w_0^2,$ $P_c\rightarrow 1$ and we obtain a directed photon emission as discussed in\cite{ref.sw1}. The appearance of an $N$ atom cooperativity parameter in the form ${\mathcal C}= N/( 2 k^2 w_0^2)$  appears to be 
generic\cite{vuletic}.

We now demonstrate a speedup in state readout using  mixed entanglement and directed emission. First consider conventional readout of a single atom qubit based on strong driving of  the closed transition between state $|a\rangle$ and the auxiliary state $|c\rangle$\cite{ref.winelandshelving}. The scattered photons are collected by a lens with solid angle $\Omega_d$ and counted by a detector with quantum efficiency $\eta.$ The number of counts obtained after time $t$ is
\begin{equation}
q_1\simeq\eta\frac{\Omega_d}{4\pi}\frac{t}{2\tau} 
\end{equation}
where $\tau$ is the radiative lifetime of the excited state. 
A measurement strategy that uses  a threshold condition in the presence of a fixed background rate results in a detection fidelity that improves exponentially with $q_1.$ Here, and in the rest of the paper, we base numerical estimates on atomic parameters corresponding to $^{87}$Rb ($\tau = 27~\rm ns$) and optical detection parameters corresponding to a possible experimental implementation\cite{ref.swqclong} ($\eta=0.6$, $\Omega_d/4\pi=.042$, and background rate of $10^4~\rm s^{-1}$). These parameters imply   $t\sim100~\mu\rm s$ for a measurement error $O(10^{-4})$.  Since this time is much longer than the nominal qubit operation time of $1~\mu\rm s$ we can use fast Rydberg operations to reduce the counting time needed. 

We  assume a qubit array with collective qubit sites interspersed with single atom logical qubits as shown in Fig. \ref{fig.qubits}. The collective qubits are trapped in an ellipsoidal volume as described above. To measure the state of $|\psi\rangle$ we perform the following steps: i) CNOT operation $|\psi\rangle|\phi\rangle\rightarrow|\psi\rangle|\phi\oplus\psi\rangle$ ,  ii) mapping of $|\phi\rangle$ to $|\phi_e\rangle$, and iii)   detection of the radiated photon. We repeat this process until the state has been measured with the desired fidelity. With $t_1$ the time needed to complete 
steps i) and ii),
and choosing the collection optics to subtend a solid angle of 
$\Omega_c,$ gives a
photon count number after time $t$ of 
\begin{equation}
q_N\sim\eta \frac{{\mathcal C}}{1+{\mathcal C}}
\frac{t}{2\tau+t_1} .
\label{eq.qn}
\end{equation}

\begin{figure}[!t]
\begin{minipage}[c]{7.5cm}
\includegraphics[width=7.5cm]{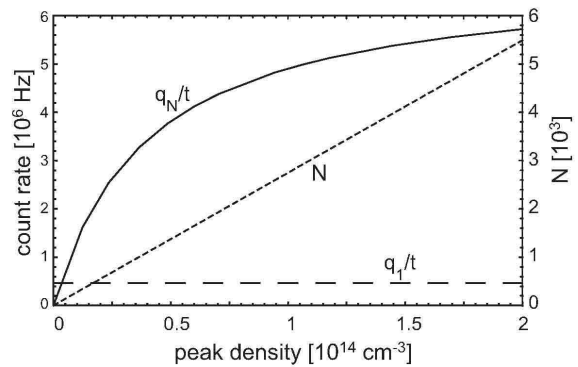}
\caption{Detected count rates from single atom qubits (dashed line) and $N$ atom qubits (solid line) as a function of the peak atomic density. The dotted line shows the number of atoms in the FORT.  Parameters used were 
$\Omega_d/4\pi=.042,$  $\eta=0.6,$ $\tau = 27 ~\rm ns,$ $t_1=33~\rm ns,$ 
$w_f=3~\mu\rm m$, $\lambda_f=1.06~\mu\rm m,$ $T_{rel}=.05,$ and $\lambda=.78~\mu\rm m,$  giving $w_a=0.82~\mu\rm m,$ $\xi=8.9$, 
 and $\Omega_c/4\pi=9.1\times 10^{-4}$.    }
\label{fig.counts}
\end{minipage}
\end{figure}

Figure \ref{fig.counts} compares the detected count rates from single atom and collective qubits. Numerical estimates show that fast coupling to the Rydberg levels can be obtained with available cw laser sources and we have assumed a CNOT rate of 30 MHz. At the peak density of $n_0=2\times 10^{14}~\rm cm^{-3}$ shown in the figure the count ratio is $q_N/q_1\sim 12$ which implies that state detection with the same fidelity could be performed 
more than 10 times faster. We note that recent experiments have demonstrated the feasibility of reaching these densities in 
a far off resonance optical trap\cite{chapmanfort,madfort}.  
The detection solid angle assumed  in Fig. \ref{fig.counts} 
corresponds to a numerical aperture of about $0.40$ $(\Omega_d/4\pi=.042)$.
 While it is possible to use larger aperture detection optics to reduce the single qubit detection time,  it is difficult to combine large optics with other experimental constraints. 
Using collective qubits for state readout enables rapid measurements while only using a small solid angle of $\Omega_c/4\pi=9.1\times 10^{-4}$
in the example given.

Several sources of decoherence must be considered in evaluating the feasibility of this approach. The first question is the fidelity of the CNOT operation that entangles $|\psi\rangle$ and $|\phi\rangle.$ This will be worse than the fidelity of a CNOT between single atom qubits due to imperfect dipole blockade and the possibility of multiple excitation of the qubit. It was shown in Ref. \cite{ref.blockade2} that the CNOT fidelity scales as $|\Omega_{\rm eff}|^2/<\Delta_{\rm dd}>^2,$ where $\Omega_{\rm eff}=\sqrt N \Omega$ is the effective $N$ atom Rabi frequency and $<\Delta_{\rm dd}>$ is the dipole-dipole interaction frequency averaged over the $N$ atoms. Referring to Fig. 1 in \cite{ref.sw1}
we see that realistic parameters can be chosen that give error probabilities 
of $O(10^{-5})$ with 500 atoms. We expect the scaling for a CNOT operation, which involves a single collective Rydberg blockade cycle, 
will be comparable. Thus with the $N\sim 5000$ atoms required in Fig. \ref{fig.counts} the error probability is a tolerable $O(10^{-4}).$
The high density inside the collective qubit results in collisional decoherence.  The density dependent collisional clock shift in $^{87}$Rb was recently measured to be\cite{clairon} $\Delta \nu/\nu\sim 30 \times 10^{-24} ~{\rm [cm^3]} \times n_0$ when the $m=0$ Zeeman levels are used.
This implies that at $n_0=2\times 10^{14}~\rm cm^{-3}$ the collisional decoherence time of $|\phi\rangle$ is about $ 4~\rm m s$. 
Additionally the decoherence time due to inelastic ground state collisions at these densities is about\cite{wiemanoverlapping} $0.1~\rm s$. 
These times are much longer than the projected time to make a measurement ($t_{\rm meas}\sim 10 ~\mu\rm s$) and can be neglected. On the other hand we note that the collective decoherence time is much shorter than the decoherence time of single atom qubits which are therefore strongly preferred for computational and storage operations.

\begin{figure}[!t]
\begin{minipage}[t]{8.cm}
\includegraphics[width=8.cm]{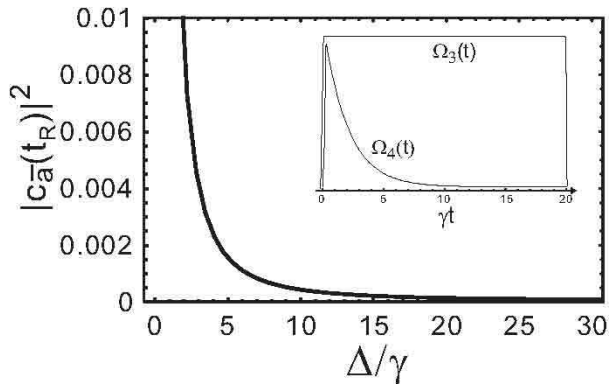}
\caption{Probability of $|\phi\rangle$ remaining in state $|\bar a\rangle$ after single photon absorption.     }
\label{fig.mappingr}
\end{minipage}
\end{figure}

\section{Quantum state transmission}
\label{sec.transmission}

As a prelude to discussing a protocol for quantum state transmission we show that the time reversed version of the above readout process can be used to detect a  coherently emitted photon by a collective receiving qubit in a target array as shown in Fig. \ref{fig.qubits}. The basic idea of converting a photonic state to an atomic excitation was introduced in Ref. \cite{ref.lukinpolariton} and discussed in \cite{ref.kwiat,ref.imamoglu} for high quantum efficiency photon detection. Consider the situation where the receiving qubit $|\phi\rangle_2$ is prepared 
in the state $|\bar a\rangle_2$ and driven by the same tri-chromatic field  used for photon generation in  Fig. \ref{fig.levels}. In the limit of a strong tri-chromatic driving field an incident photon has probability $P_0$ of not being absorbed and probability $P_1=1-P_0$ of being absorbed by the receiving qubit. If the receiver is prepared in the state $|\bar a\rangle$ absorption of a photon will result in the transformation

\begin{equation}
|1_{\bf k}\rangle|\bar a\rangle  \rightarrow 
\sqrt P_1|0_{\bf k}\rangle \frac{1}{\sqrt N}\sum_j e^{\imath\varphi_j} |a_{\tilde \jmath};b_j\rangle
\label{eq.mappingr}
\end{equation}
with $\varphi_j=(-{\bf k}-{\bf k}_3+{\bf k}_2+{\bf k}_1)\cdot{\bf r}_j.$ 
When ${\bf k}={\bf k}_4= {\bf k}_1+{\bf k}_2-{\bf k}_3$ the process is phase matched and we generate the singly excited state $|\phi\rangle_2=(1/\sqrt N)\sum_j  |a_{\tilde \jmath};b_j\rangle$ with a probability collectively enhanced by a factor of  $N.$ 
For a cigar shaped qubit as described above $P_1\simeq 1-e^{-N\sigma/A}$ with 
$\sigma = 3\lambda^2 /(2\pi)$ and $A=\pi w_0^2.$ 
With $\lambda=0.78~\mu\rm m$ and $w_0=2~\mu\rm m$ only five hundred  atoms gives $P_1=1-10^{-5}.$  
Note that, as discussed in Sec. \ref{sec.readout} for state readout, the tri-chromatic field could also be replaced with a single field connecting levels $|b\rangle$ and $|c\rangle$ in Fig. \ref{fig.levels}. The use of a tri-chromatic field may have advantages from the perspective of using  the ensemble qubits interchangeably both as transmitters and receivers.

\begin{figure}[!t]
\begin{minipage}[t]{7.5cm}
\includegraphics[width=7.cm]{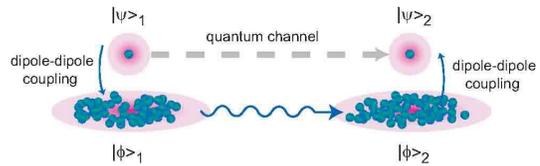}
\caption{(color online) A quantum channel between $|\psi\rangle_1$ and $|\psi\rangle_2$ is created by dipole-dipole coupling to ensembles that exchange photons.     }
\label{fig.atomcoupling}
\end{minipage}
\end{figure}

To ensure the mapping of Eq. (\ref{eq.mappingr}) is achieved with high probability several conditions must be satisfied. Multi-photon resonance for the Raman process requires 
$\Delta+\Delta_{ac}=0$ where 
$\Delta=(\omega_1+\omega_2-\omega_3-\omega_4)-\omega_{ab}$ is the four-photon detuning and $\Delta_{ac}$ accounts for AC Stark shifts of the levels.  To avoid spontaneous emission from the excited levels all single photon  transitions should be strongly detuned, and the tri-chromatic dressing field should be gated on in a time window coincident with the known arrival time of the photon\cite{ref.kimblerepeater} to avoid rotating the atom transferred to $|b\rangle$ back to the upper level $|c\rangle,$ from where there is a finite probability for unwanted decay down to $|a\rangle.$

To estimate the fidelity of the photon detection process we define an effective Rabi frequency for the three photon dressing field $\Omega_3(t)$ and model the single photon pulse as $\Omega_4(t)=
\Omega_{40}e^{-\gamma(t-t_4)/2}$ with $\gamma$ the radiative decay rate of state $|c\rangle$ and $t_4$ the random emission time of the photon.
The single photon pulse can at most contribute to the transfer of one atom from $|a\rangle$ to $|b\rangle.$ Assuming four photon resonance and adiabatically eliminating the excited states which have small populations, the Schr\"odinger equation for the slowly varying amplitudes of 
$|\phi\rangle=c_{\bar a}|\bar a\rangle + c_{\bar b}|\bar b\rangle$
gives 
\begin{subequations}
\begin{eqnarray}
 \frac{dc_{\bar a}}{dt}& =&  -i\sqrt N \frac{\Omega_R(t)}{2(1+i\gamma/2\Delta)}c_{\bar b}  \\
 \frac{dc_{\bar b}}{dt} &= & -i\sqrt N \frac{\Omega_R^*(t)}{2(1+i\gamma/2\Delta)}c_{\bar a}
\end{eqnarray}
\label{eqs.6}
\end{subequations}
\noindent where the effective Rabi frequency is
$\Omega_R(t)=\Omega_3(t)\Omega_4^*(t)/2\Delta$ with $\Delta=\omega_4-\omega_{ca}$ the excited state detuning.   
These equations are derived using an effective non-Hermitian Hamiltonian and do not conserve probability. They 
are useful as long as the excited state populations are small. 
As shown in Fig. \ref{fig.mappingr} we gate the dressing field on with a fixed amplitude $\Omega_{30}$ for a window of length $t_R=20/\gamma.$ With the Rabi frequencies normalized such that $\sqrt N\int_0^{t_R} dt ~\Omega_R(t)=\pi$ we find that the probability of transfer failure is less than $O(10^{-3})$ for $\Delta/\gamma>20.$ 
For a single photon pulse in a spatial mode with waist of $w_0=3~\mu\rm m$ and using parameters corresponding to the $^{87}$Rb cycling transition the peak Rabi frequency is $\Omega_{4,\rm max}\simeq 2\pi\times 0.9~\rm MHz.$ Using the $\pi$ pulse normalization of $\Omega_R$ given above with $t_R=20/\gamma$ we find 
$\Omega_3\sim \pi \Delta\gamma /(10 \sqrt N \Omega_{4,\rm max})\sim 2\pi \times  0.25  ~\rm MHz$ for $\Delta=20 \gamma$ and $N=10^3.$  The corresponding excited state populations are approximately $|\Omega_{4,\rm max}/\Delta|^2 $ and 
 $|\Omega_{3}/\Delta|^2 $ which are both less than $10^{-4}.$ This  justifies adiabatically eliminating the excited state population in the  derivation of  Eqs. (\ref{eqs.6}).

While Fig. \ref{fig.mappingr} shows we need $\Delta\gg\gamma$ for high fidelity mapping of the photon to an atomic excitation  the photon being detected is emitted on the same $|c\rangle\to|a\rangle$ transition  that participates in the detection process. Thus we will naturally have $|\Delta|\stackrel{<}{\sim} \gamma.$ Larger $\Delta$ and high detection fidelity can be achieved by using an ac Stark shift from an  auxiliary off resonant laser to shift the transition frequency of the atoms in the receiving site.

Having established the possibility of high fidelity bidirectional transmission between collective qubits we 
can construct a photonic channel for quantum state transmission and distant entanglement between single atom qubits. The steps involved in constructing the quantum channel are shown in Fig. \ref{fig.atomcoupling}. 
To transmit a single atom qubit state $|\psi\rangle_1$ in array $1$ to a single atom qubit state $|\psi\rangle_2$ in array $2$ we initialize $|\phi\rangle_1,|\phi\rangle_2,$ to $|\bar a\rangle$ and $ |\psi\rangle_2$ to $|a\rangle$  and then perform the following sequence: 
i) CNOT operation $|\psi\rangle_1|\phi\rangle_1\rightarrow|\psi\rangle_1|\phi\oplus\psi\rangle_1$,
i')  CNOT operation $|\psi\rangle_1|\phi\rangle_1\rightarrow|\psi\oplus\phi\rangle_1|\phi\rangle_1$, 
ii) $|\phi\rangle_1$ generates a photon in mode ${\bf k}_4$, iii) mode ${\bf k}_4$ is detected by $|\phi\rangle_2$,
iv) CNOT operation $|\psi\rangle_2|\phi\rangle_2\rightarrow|\psi\oplus\phi\rangle_2|\phi\rangle_2$. This protocol gives the mappings $|a\rangle_1|a\rangle_2\rightarrow |a\rangle_1|a\rangle_2$ and $|b\rangle_1|a\rangle_2\rightarrow |a\rangle_1|b\rangle_2$. We therefore obtain 
\begin{eqnarray}
|\psi\rangle_1|a\rangle_2=(c_a|a\rangle_1 + c_b|b\rangle_1)|a\rangle_2\rightarrow \nonumber\\
|a\rangle_1(c_a|a\rangle_2 + c_b|b\rangle_2)=|a\rangle_1|\psi\rangle_2
\end{eqnarray}
which is an ideal quantum transmission\cite{cirac}. Note that if the disentangling step i') is neglected we obtain the mapping 
\begin{equation}
|\psi\rangle_1|a\rangle_2=(c_a|a\rangle_1 + c_b|b\rangle_1)|a\rangle_2\rightarrow 
c_a|a\rangle_1|a\rangle_2 + c_b|b\rangle_1|b\rangle_2.
\end{equation}
When $c_a=c_b=1/\sqrt2$ we create the Bell state $(|a\rangle_1|a\rangle_2 + |b\rangle_1|b\rangle_2)/\sqrt2$ which can be used with local measurements for teleportation of the quantum state\cite{bennettteleport}.

\begin{figure}[!t]
\begin{minipage}[t]{7.5cm}
\includegraphics[width=7.cm]{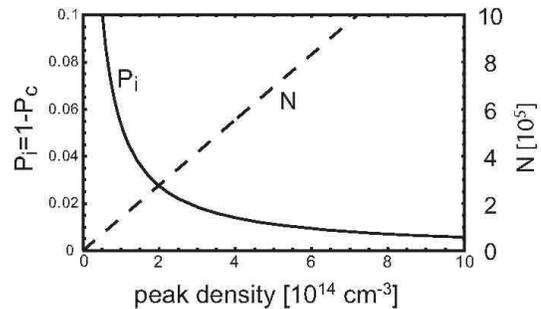}
\caption{Probability of incoherent photon emission (solid line) and number of atoms (dashed line) in the collective qubit. Parameters were 
$w_f=8~\mu\rm m$, $\lambda_f=1.06~\mu\rm m,$ $T_{rel}=.05,$ $\lambda=.78~\mu\rm m,$  giving $w_a=2.2~\mu\rm m,$ $\xi=24.$, 
 and $\Omega_c/4\pi=1.3\times 10^{-4}$.    }
\label{fig.pcoherent}
\end{minipage}
\end{figure}

Provided that local operations are performed with high fidelity the dominant source of error in this transmission protocol is the probability $P_i$
of emission of a photon in a random direction in step ii). This error can be minimized by maximizing the cooperativity parameter.
Looking at Eq. (\ref{eq.pc}) this 
implies that $k^2 w_0^2/N$ should be small, or that the product of the density times the length of the trapped cloud (i.e. the optical depth) should be maximized. 
We show in Fig. \ref{fig.pcoherent} the random emission probability as a function of peak density for optimized geometrical cloud parameters.
At densities of $5 \times 10^{14}~\rm cm^{-3}$ and $N\simeq 6\times 10^5$ we achieve $1-P_c\simeq .01.$ It should be noted that the collective qubit for these parameters occupies a larger volume than that used for fast  state detection, which implies that the dipole-dipole interaction entangling the cloud will be weaker, so that entangling operations must be performed more slowly to preserve fidelity. 

\section{Discussion}
\label{sec.discussion}
 
We have described protocols for fast quantum state detection and transmission 
based on entanglement of single atom and collective qubits. 
Enhancement in readout speeds relies on the ability to entangle a single atom with a tightly confined cloud of atoms through the long range dipole-dipole interaction of Rydberg states. Reduction in the readout time by a factor of ten appears feasible with clouds of a few thousand atoms. Techniques for loading single atom optical traps positioned in the proximity of larger many atom traps remain to be developed. One possibility may be to 
load all sites with $N$ atoms and then select a single atom in sites to be used for computation with, for example, dipole blockade ideas presented in our earlier work\cite{ref.sw1}. 

Using cross entanglement of single  atom and mesoscopic qubits for both transmitting and detecting photons leads to the possibility of establishing a quantum channel. The leading factor limiting the fidelity of such a channel is the probability for incoherent emission of a photon in a random direction. Reducing the error probability to the level of a few percent implies $N\sim 10^5$ atoms in the mesoscopic qubit.  In light of density limitations this in turn implies the mesoscopic qubit should be confined to a region about $20~\mu\rm m$ long.  Detailed calculations of Rydberg atom interactions suggest that at principal quantum number $n=95$ about 25 MHz of dipole-dipole interaction is available at a separation of $20~\mu\rm m$\cite{ref.swqclong}.
Since the speed of local operations should be small compared to the dipole-dipole interaction frequency for high fidelity operation (the relevant contribution to the gate error scales as the square of Rabi frequency/dipole-dipole frequency) a quantum channel with transmission rates as high as a MHz appear possible. 
While experimental demonstration of these ideas will certainly be challenging we present them as an  alternative to experiments using strong coupling of single atoms and single photons in  high finesse cavities.

M. S. is grateful to V. Vuletic for emphasizing the scaling of Eq. (\ref{eq.pc}).
Support was provided  by  the U. S.
Army Research Office under contract number DAAD19-02-1-0083,
and NSF grants EIA-0218341 and PHY-0205236.

\end{document}